RESEARCH AND EDUCATION

# Fatigue performance of prosthetic screws used in dental implant restorations: Rolled versus cut threads

Mikel Armentia, MSc,[a] Mikel Abasolo, PhD,[b] Ibai Coria, PhD,[c] Joseba Albizuri, PhD,[d] and Josu Aguirrebeitia, PhD[e]

## ABSTRACT

**Statement of problem.** Cold rolling is widely used for screw thread manufacturing in industry but is less common in implant dentistry, where cutting is the preferred manufacturing method.

**Purpose.** The purpose of this in vitro study was to compare the surface finish and mechanical performance of a specific model of prosthetic screw used for direct restorations manufactured by thread rolling and cutting.

**Material and methods.** The thread profiles were measured in an optical measuring machine, the residual stresses in an X-ray diffractometer, the surface finish in a scanning electron microscope, and then fatigue and static load tests were carried out in a direct stress test bench according to the International Organization for Standardization (ISO) 14801. Finally, linear regression models and 95% interval confidence bands were calculated and compared through ANCOVA for fatigue tests while the *t* test was used for statistical comparisons ($\alpha$=.05).

**Results.** The surface finish was smoother, and compressive residual stresses were higher for the roll-threaded screws. Linear regression models showed a fatigue life 9 times higher for roll-threaded screws (*P*=1) without affecting static behavior, which showed statistically similar static strengths (*P*=.54). However, the thread profile in the roll-threaded screws was not accurately reproduced, but this should be easily corrected in future prototypes.

**Conclusions.** Rolling was demonstrated to be a better thread-manufacturing process for prosthetic screws, producing improved surface quality and fatigue behavior. (J Prosthet Dent 2021;126:406.e1-e8)

Dental implants are widely used to replace missing teeth.[1] In direct-to-implant restorations, implant and abutment are joined by the preload obtained through the tightening torque applied to the prosthetic screw[2] as illustrated in Figure 1, providing the necessary structural integrity to the restoration.[3-8] The recommended torque value is provided by manufacturers based on different implant design factors.[9]

Clinical studies suggest that the failure of implant restorations is often induced by fatigue because of the variable load conditions during their life span.[10-12] These failures typically occur in the prosthetic screw, with abutment or implant failures being less common.[13,14] The screw fracture is usually located at the first thread engaged with the implant.[15,16] Under given loading conditions, the number of cycles to fatigue failure depends on several parameters, including component dimensions, screw metric, material, manufacturing processes, and preload level.[17] Most prosthetic screws are made of pure or alloyed titanium, as these are less expensive than gold alloy screws while having excellent biocompatibility, corrosion resistance, machinability, and desirable physical and mechanical properties.[18,19] Two processes are widely used for screw thread manufacturing in industry: cutting and cold rolling. In thread cutting, material is removed from a cylindrical blank by machining, and in thread rolling, a matched set of dies displaces material of the manufacturing part to produce external threads on the

Supported by the Basque Government (Grant number IT947-16).
[a]Doctoral student, Department of Mechanical Engineering, University of the Basque Country, Bilbao, Spain; and R&D Engineer, Biotechnology Institute I mas D S.L., Miñano, Spain.
[b]Associate Professor, Department of Mechanical Engineering, University of the Basque Country, Bilbao, Spain.
[c]Lecturer and Researcher, Department of Mechanical Engineering, University of the Basque Country, Bilbao, Spain.
[d]Associate Professor, Department of Mechanical Engineering, University of the Basque Country, Bilbao, Spain.
[e]Associate Professor, Department of Mechanical Engineering, University of the Basque Country, Bilbao, Spain.

## Clinical Implications

Thread rolling is a suitable manufacturing process for prosthetic screw threads that significantly enhances the fatigue performance of dental restorations.

cylindrical blank in a cold forming operation with no material losses.[20]

Thread rolling offers many benefits in terms of manufacturing and mechanical behavior. The rolling process is less time-consuming because the thread may be obtained in a single pass[21,22] with no need for secondary operations,[23] which significantly increases productivity[24] and reduces unit product cost.[21] Regarding mechanical behavior, the thread rolling process introduces compressive residual stresses[25] on the thread surface, increasing hardness from strain hardening.[26] Furthermore, as no material is removed, good grain flow is obtained,[21,27] improving surface quality.[22,28] Consequently, rolled threads have been reported to provide better fatigue results than cut threads.[29-31]

Even though rolled threads are widely used in general industry,[21,24,29] they are less common in threaded prosthetic components, possibly because of the small screw sizes. The purpose of the present work was to compare both manufacturing techniques in a direct restoration prosthetic screw in terms of thread profile, residual stresses, surface finish, and fatigue and static mechanical behavior. The hypothesis of this study was that rolled threads would provide better mechanical response than cut threads.

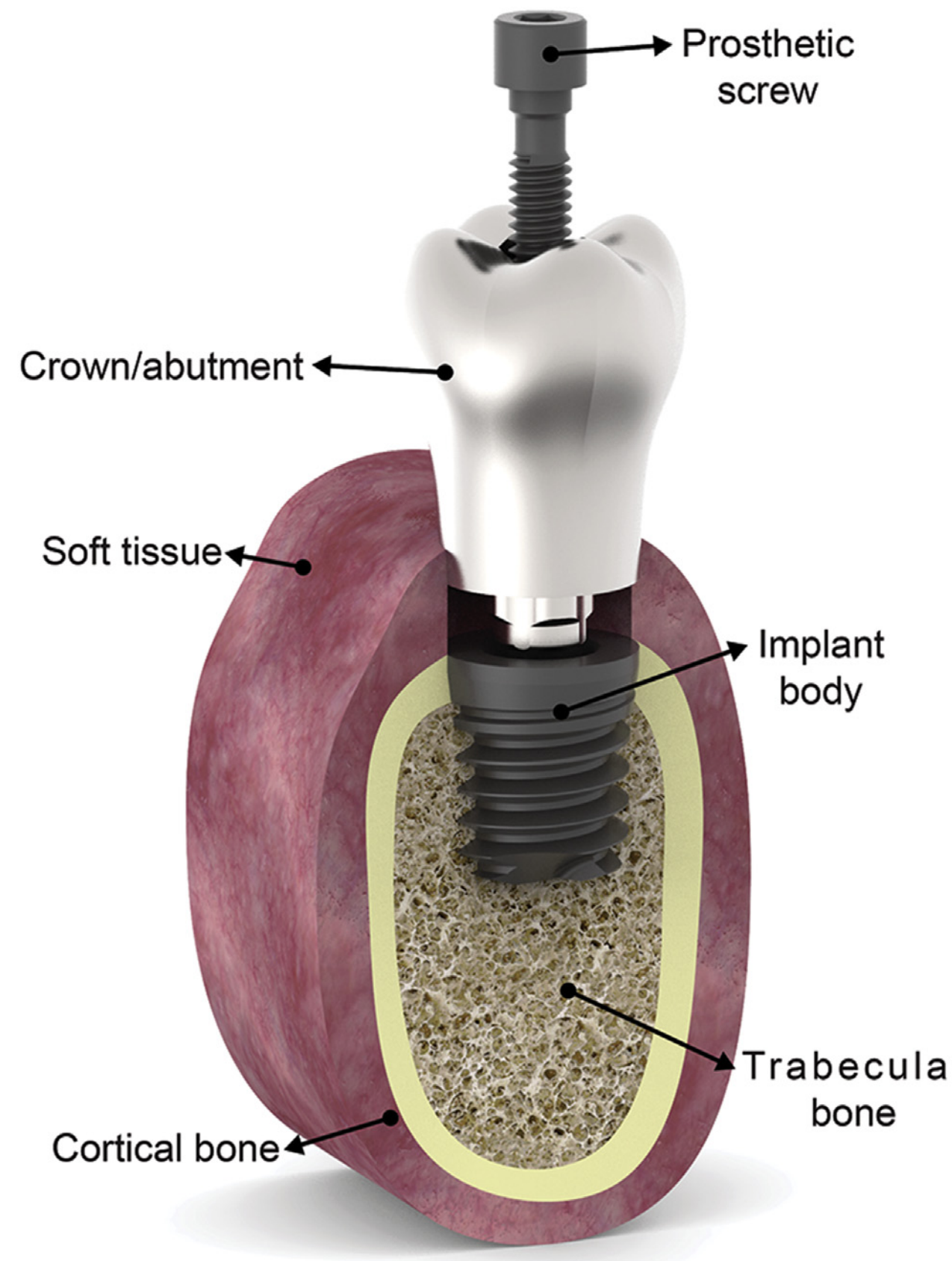

**Figure 1.** Assembly of single dental implant restoration.[17]

## MATERIAL AND METHODS

The retaining screw studied (INTTUH; BTI Biotechnology Institute) has been commonly used by the dental implant manufacturer BTI (BTI Biotechnology Institute), with a 1.8-mm metric thread and with a tungsten carbide surface treatment (Ti Black; BTI Biotechnology Institute) to reduce the friction coefficient and maximize the preload for a given tightening torque. Two batches of INTTUH prosthetic screws were manufactured: one with cut threads and the other with rolled threads. Most machining operations were the same for both batches, except for the diameter of blank material before threading, the turning speed, the feed rate, and the number of passes during the thread-manufacturing process. For the cut threads, a Ø1.8-mm blank was used, and 20 passes were made at 5000 rpm with a 1750-mm/min feed rate. For the rolled threads, a Ø1.5-mm blank was used, and the final profile was obtained by 1 pass at 200 rpm with a 70-mm/min feed rate. In both operations, the same commercially available lubricant was used (Blasomill 10; Blaser Swisslube). For the measurements of the thread profile, surface roughness, and compressive residual stress, the coating was not added. Then, the prosthetic screw was mounted in an implant (IIPSCA4513; BTI Biotechnology Institute) with a Ø4.5-mm body, a Ø4.1-mm universal platform, and a titanium abutment (INPPTU44; BTI Biotechnology Institute), as used for directly attached implant-supported restorations, all provided by the manufacturer. The static and fatigue failure tests were carried out with a coated (Ti Black; BTI Biotechnology Institute) screw. Figure 2 shows the assembly where, following the recommendation of the manufacturer, a tightening torque of 35 Ncm was applied to the screw. The retaining screw was made of Ti 6Al 4V ELI (Ti GR5), while the implant and the abutment were made of Ti CP4, the chemical composition of which is provided in Table 1.

Cutting and rolling operations may cause slight dimensional differences in the thread profile. Therefore, a multisensor measuring machine (Zeiss O-Inspect 322; Carl Zeiss Iberia, S.L.) was used to examine both thread profiles and measure the thread parameters, including the internal diameter, external diameter, pitch, and thread angle for 1 specimen of each batch of thread rolled and cut screws.

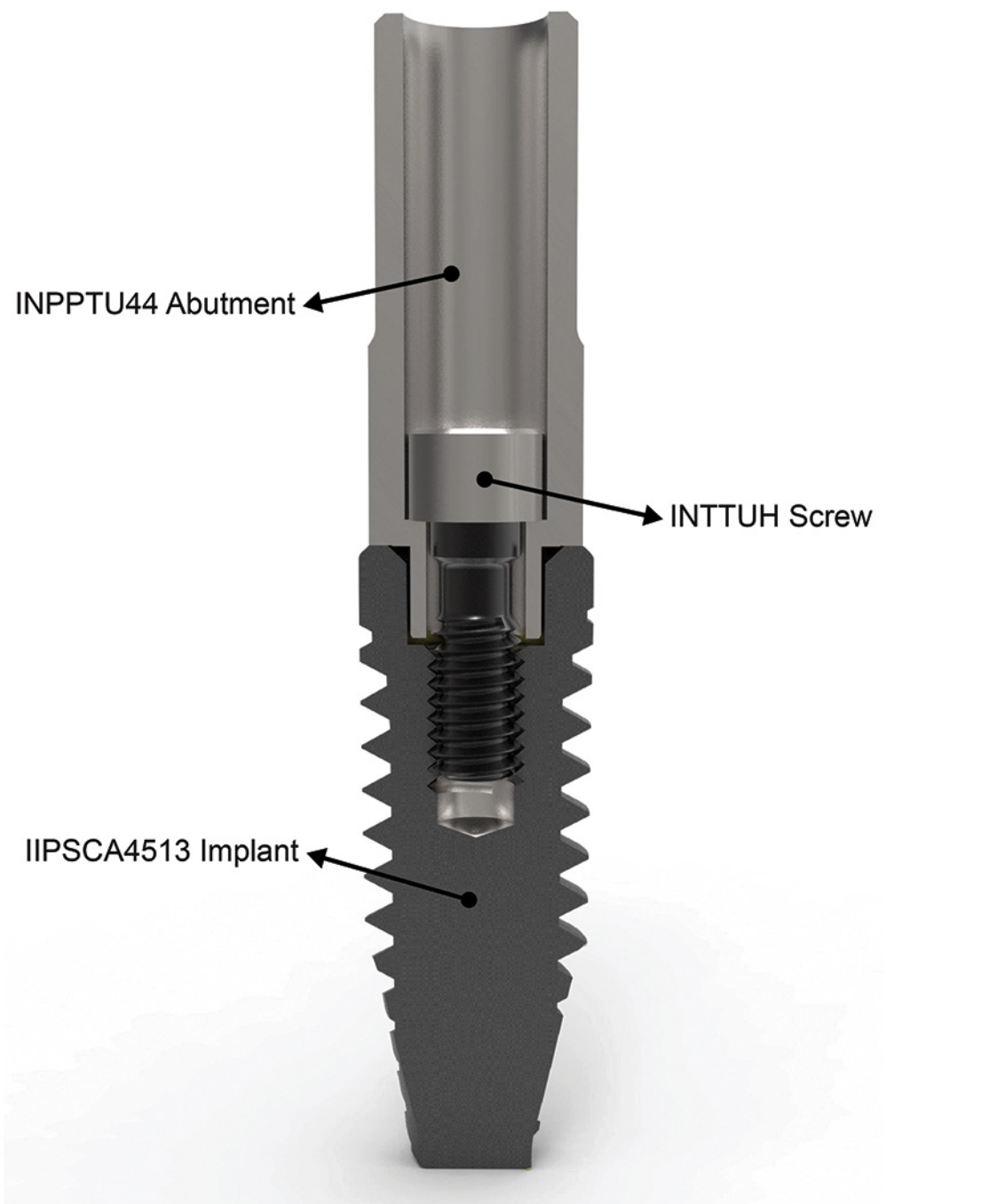

Figure 2. Dental implant assembly: implant body, prosthetic screw, and abutment.[17]

The surface roughness of 1 specimen of each batch was measured with a scanning electron microscope (Thermo Scientific Phenom ProX; ThermoFisher Scientific/Paralab). Surface analyses were performed for a field of view of 53.7 μm, and the results were provided after Gaussian filtering (microroughness filter $\lambda_s$=20 nm and waviness filter $\lambda_c$=20 μm).

The residual compressive stresses in the screw thread surface were measured in 3 specimens for each manufacturing technique (thread rolling and cutting) by using a diffractometer (Bruker D8 Discover; Bruker) equipped with a Chromium Point or Line Focus X-ray tube, V filter (λ=2.2911 Å), PolyCapTM (1-μm single crystal cylinders) system for parallel beam generation (divergence of 0.25 degrees), and a 1-D detector (LynxEye; Bruker) with active length in 2θ 2.7 degrees. The twist X-ray tubes allowed for the quick selection change between point and line focus.[32-34] The prosthetic screws were mounted on an Eulerian Cradle with an automatically controlled X-Y-Z stage. Data were collected from 59 to 64.2 degrees 2θ (step size=0.05 and time per step=7.5 seconds). Strain values in side inclination mode were recorded for different specimen tilt angles (0, 18.4, 26.6, 33.2, 39.2, 45.0, 50.8, and 56.8 degrees) at constant azimuth angles phi. Strain-Sin2 ψ was plotted to estimate the stress values. In order to acquire a complete evaluation, at least 6 measurements were needed on the Strain-Sin2 ψ plot by using 3 different values of phi, and 0, 45, and 90 degrees were chosen in negative and positive values. Stress was evaluated from strain values by using

Table 1. Chemical composition of materials used in implant and prosthetic components

| Ti 6Al 4V ELI (Ti GR5) | | Ti CP4 | |
|---|---|---|---|
| Composition | Wt. % | Composition | Wt. % |
| Al | 5.5–6.5 | N(max) | 0.05 |
| V | 3.5–4.5 | C(max) | 0.08 |
| Fe(max) | 0.25 | Fe(max) | 0.5 |
| O(max) | 0.13 | O(max) | 0.4 |
| C(max) | 0.08 | H(max) | 0.0125 |
| N(max) | 0.05 | - | - |
| H(max) | 0.012 | - | - |

Wt. %, weight percentage.

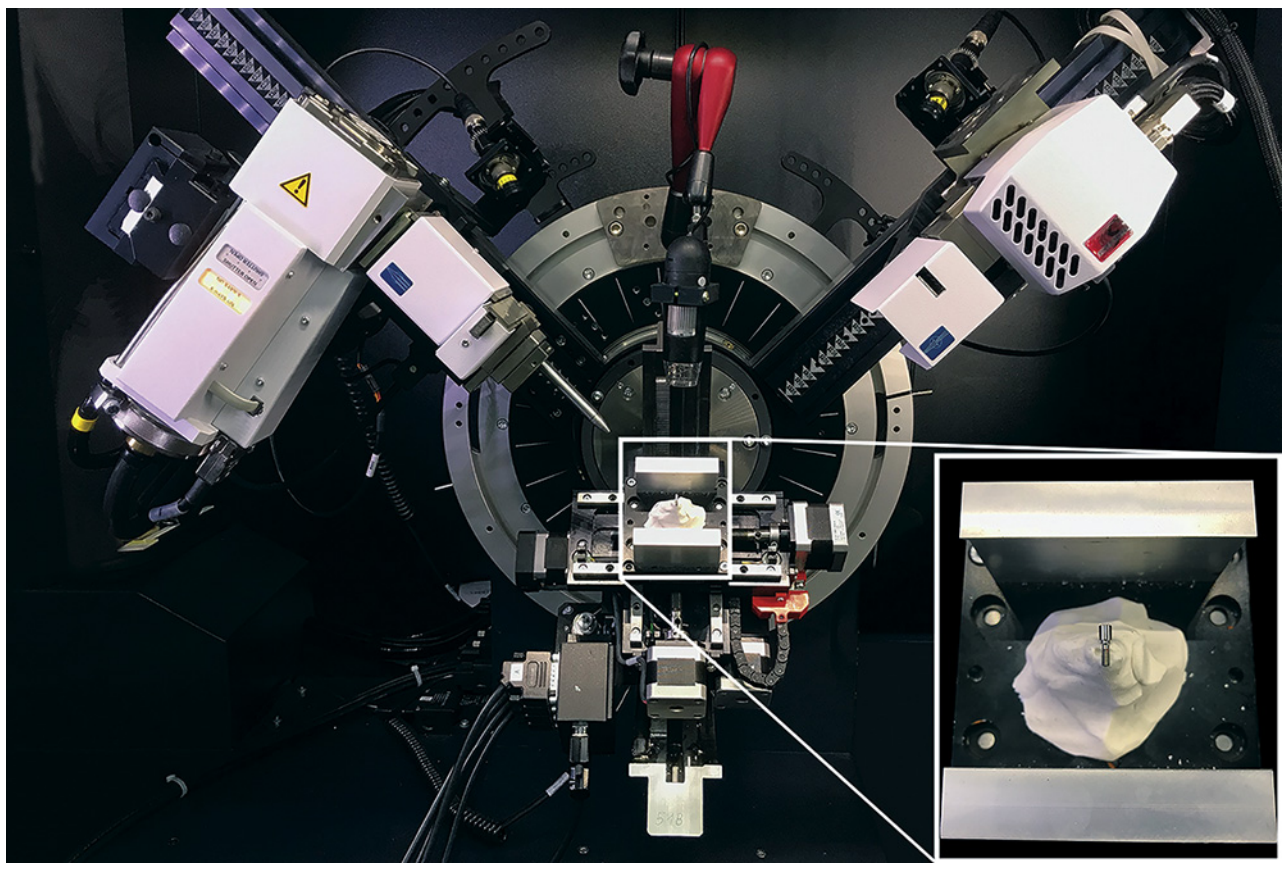

Figure 3. Diffractometer: general view and detail of specimen.

the Young modulus E (111 982 MPa) and Poisson ratio ν (0.330) and by taking into consideration the elastic constants s1 ($-2.947\times10^{-6}$) and ½ s2 ($1.188\times10^{-5}$) of the material. A single peak (101), available at 61.4 value of 2θ, was used for the analysis. The obtained results were adjusted by using a software program (Leptos 7.03; Bruker AXS GmbH). The data were corrected for absorption, background (5 points at edges), polarization, smoothness, and K alpha2 subtraction, and the peak evaluation was fitted by the Pearson VII function. The determined values were obtained by using a biaxial mode with the Psi splitting function because of the shear stress components. Figure 3 shows the experimental design.

The dental implant assembly (Fig. 2) was tested under static and cyclic loading conditions to evaluate its mechanical strength. A direct stress fatigue test bench (E 3000 Electropuls; Instron) with a ±5-kN load range load cell (DYNACELL 2527-153; Instron) was used. Tests were carried out in accordance with the International Organization for Standardization (ISO) 14801 Standard,[35] which established the test conditions for dental implant fatigue tests. In brief, the implant was fixed in a rigid clamping device with the implant extending 3 mm from the specimen holder and with the load applied at 8 mm from the implant-abutment connection with an inclination of 30 degrees with respect to the implant axis.

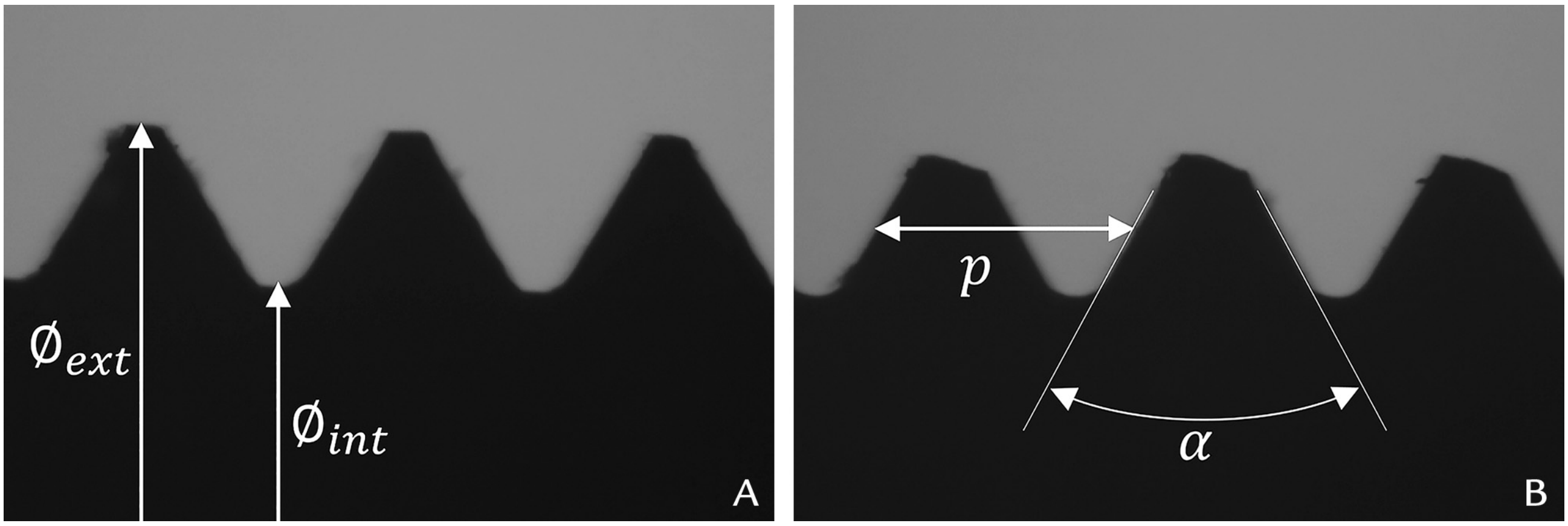

**Figure 4.** Thread profile. A, Cut thread. B, Rolled thread. Original magnification ×150.

For fatigue tests, the load varied sinusoidally between the test load and 10% of this value (R0.1) with a frequency not higher than 15 Hz, and the number of cycles to failure was reported. Thus, the curve that related the applied force with the number of cycles was obtained. For such purpose, a total of 24 fatigue tests were performed fulfilling ISO 14801 requirements: 9 tests at 3 different load levels (3 specimens per load level) for cut threads and 15 tests at 3 different load levels (5 specimens per load level) for rolled threads because a longer fatigue life and consequently more scatter were expected.[36] The cut thread tests had been previously published.[17] However, both tests were carried out at the same fatigue tests bench and in the same period, making the results comparable. For the static tests, the load was quasistatically incremented until collapse, and the collapse load was reported as a result. Five specimens were tested for both cut and rolled threads. For the statistical analysis of the results, ANCOVA was used to compare linear regression models from fatigue data, and a *t* test ($\alpha$=.05) was used to compare static strength.

## RESULTS

Figure 4 shows the profiles of the rolled and cut threads. Table 2 presents the dimensions obtained from both manufacturing processes: the external diameter $\varnothing_{ext}$, the internal diameter $\varnothing_{int}$, the pitch $p$, and the angle of the thread profile $\alpha$. Unlike cut threads, where thread crests were parallel to the longitudinal axis of the screw, rolled thread crests showed a different shape. Thus, 2 values of the external diameter were measured in this case: maximum and minimum. The equivalent stress area $A_s$, which is the effective resistance area for threaded sections,[37] was calculated as in a previous study.[17] Cut threads showed an equivalent tensile stress area of 1.37 mm$^2$, while rolled threads showed 1.50 mm$^2$.

Figure 5 shows the surface roughness obtained from both manufacturing processes. The cutting process removed material, leaving transverse lines from the irregular shape of the turning insert caused by wear and the several passes used in this manufacturing process. In contrast, material flow was seen with the rolling process, indicating that material was formed rather than removed. Table 3 gives the results of the surface roughness measurements: Area roughness average $S_a$ was obtained for the whole field of view, while roughness average $R_a$ and mean roughness depth $R_z$ of a path perpendicular to the cutting or rolling direction were calculated.

**Table 2.** Dimensions of thread profile (N=1)

| Thread Type | $\varnothing_{ext}$ (mm) | $\varnothing_{int}$ (mm) | Pitch (mm) | $\alpha$ (degrees) |
|---|---|---|---|---|
| Cut thread | 1.74 | 1.24 | 0.35 | 60.5 |
| Rolled thread | 1.69/1.60 | 1.30 | 0.35 | 58.8 |

$\alpha$, angle of the thread profile; $\varnothing_{ext}$, external diameter; $\varnothing_{int}$, internal diameter.

Table 4 shows the compressive residual stresses in the longitudinal direction of the screw. The specimen geometry (screw thread) and the material absorption generated high standard errors in the measurements, a consequence of a high dispersion of the measured data. Nevertheless, the measurements confirmed, at least qualitatively, that thread rolling introduced considerably higher compressive loads than thread cutting.

As a result of the fatigue tests, a curve was obtained that related the applied cyclic load magnitude with the number of cycles to fatigue failure. The curve for a dental implant with cut-threaded screw has been previously published in a study[17] where 9 experimental tests were performed at 3 different load levels: 325, 350, and 375 N. In the present work, the same tests were carried out with roll-threaded screws. For this experiment, 15 specimens were tested (5 tests at each load level) because larger fatigue life and, consequently, more scatter were expected.[36] Figure 6 shows the experimental results along with the linear models log(F)-log(cycles) and 95% confidence bands according to the ASTM E-739 Standard.[38] Even though linear regression models were obtained

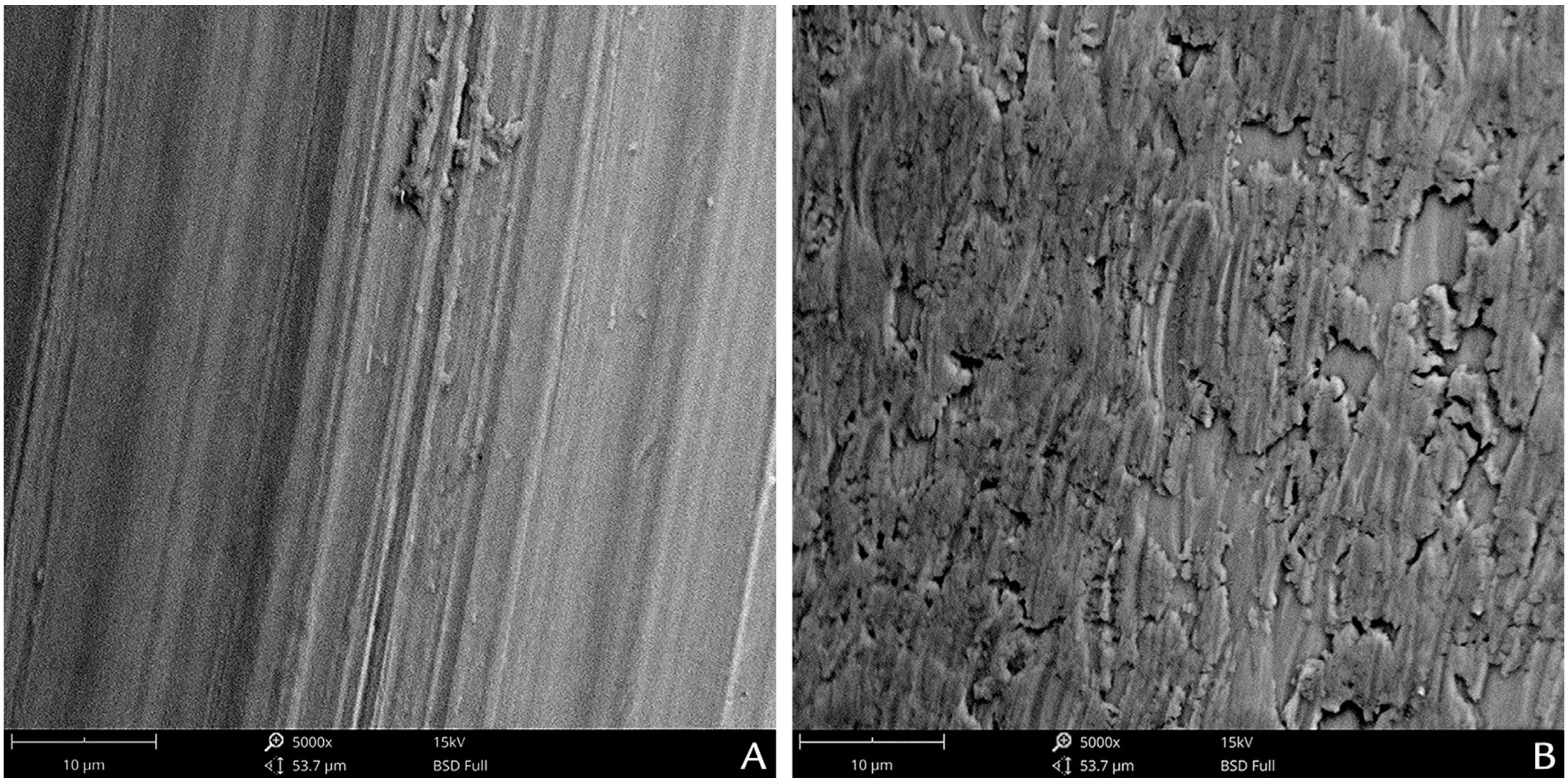

**Figure 5.** Surface roughness measurements. A, Cut thread. B, Rolled thread. Original magnification ×5000.

**Table 3.** Surface roughness parameters (N=1)

| Thread Type | $S_a$ (nm) | $R_a$ (nm) | $R_z$ (µm) |
|---|---|---|---|
| Cut thread | 430 | 446 | 1.95 |
| Rolled thread | 333 | 434 | 1.83 |

$R_a$, roughness average; $R_z$, mean roughness depth; $S_a$, area roughness average.

**Table 4.** Mean values and standard errors of compressive residual stresses in threaded area

| | Residual Stresses (MPa) | |
|---|---|---|
| N=48 (per Test) | Cut Thread | Rolled Thread |
| Test 1 | 281.5 ±136.1 | 569.5 ±87.7 |
| Test 2 | 264.9 ±119.3 | 480.0 ±103.6 |
| Test 3 | 297.0 ±124.8 | 496.0 ±87.2 |

from logarithmic values for both force and cycles, only the horizontal axis (cycles) is shown in log scale in Figure 6 to facilitate interpretation of the results. While the failure section in cut screws was in the first engaged thread, the failures in the rolled screws took place in the head-shank transition section (Fig. 7). ANCOVA was used to compare both linear models, accepting the first hypothesis that the slopes were equal ($P$=1) and rejecting the second hypothesis ($P$<.001); that is, the mean fatigue life was statistically different. Furthermore, once the slopes of both models had been determined to be equal, the fatigue life was calculated to be 9 times larger for roll-threaded screws ($P$=1).

Table 5 shows the maximum static loads before fracture of the dental implant assembly. In this case, the failure in both cut and rolled threads took place in the first engaged thread. A *t* test was carried out for the static loads registered in Table 5. The null hypothesis was accepted ($P$=.54); that is, the static load was assumed to be statistically the same for both manufacturing processes.

## DISCUSSION

The hypothesis of the study was accepted as the results indicated that rolled threads had better mechanical behavior than cut threads. Even though the obtained geometries were similar and functional, the crests of the rolled thread poorly reproduced the desired geometry. This shape was obtained because of grain flow caused by the rolling process.[39,40] The small diameter of the blank may have increased the effect of this phenomenon, so the diameter of the blank should be optimized in future studies.

The rolled threads had lower surface roughness parameter values, that is, better surface finish and quality, consistent with previous studies.[21,28,41,42] The surface finish was mainly influenced by the residual compressive stresses on the thread surface by the rolling process, found to be larger than the ones left by cutting.[25,31] Using synchrotron radiation to evaluate surface properties should improve the quality of the data and allow the extrapolation of stress more accurately.[43-45]

The prosthetic screws had a significantly better fatigue response with thread rolling, consistent with roll-threaded screws used in industry.[29-31] Linear regression models showed a fatigue life 9 times higher for roll-threaded screws. In addition, roll-threaded screws failed at the head-shank transition section (Fig. 7) rather than at the first engaged thread, suggesting an increased improvement in the threaded section. Presumably, the improved fatigue behavior of rolled threads was mainly because of the larger compressive residual stresses because differences in the surface roughness and tensile stress area were small. The static load at failure was statistically similar for both manufacturing processes.

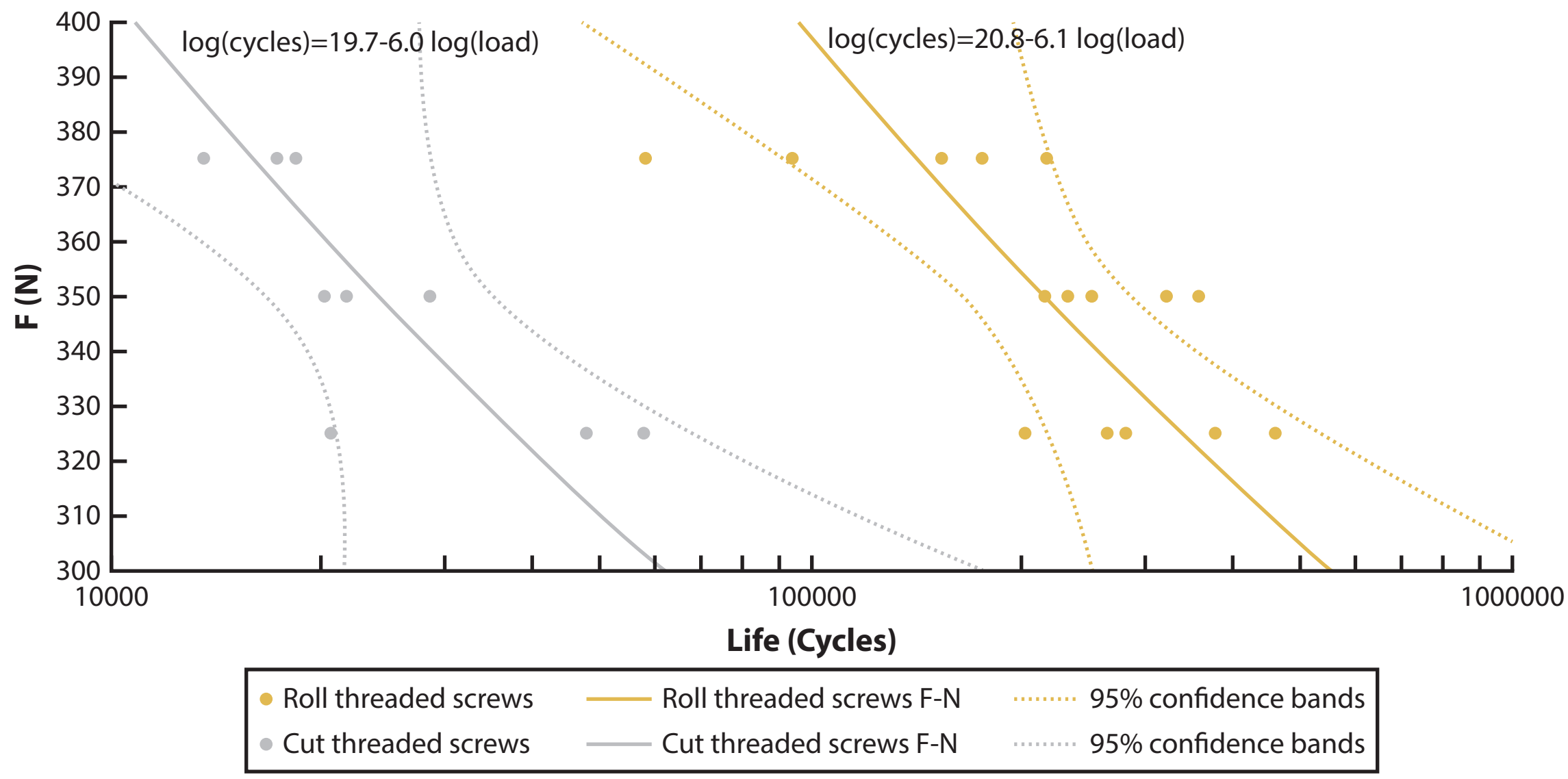

**Figure 6.** Fatigue data and linear models (F-N curves) for dental implant assemblies mounting cut- and roll-threaded screws. Data points and linear model for cut threads extracted from previous study.[17]

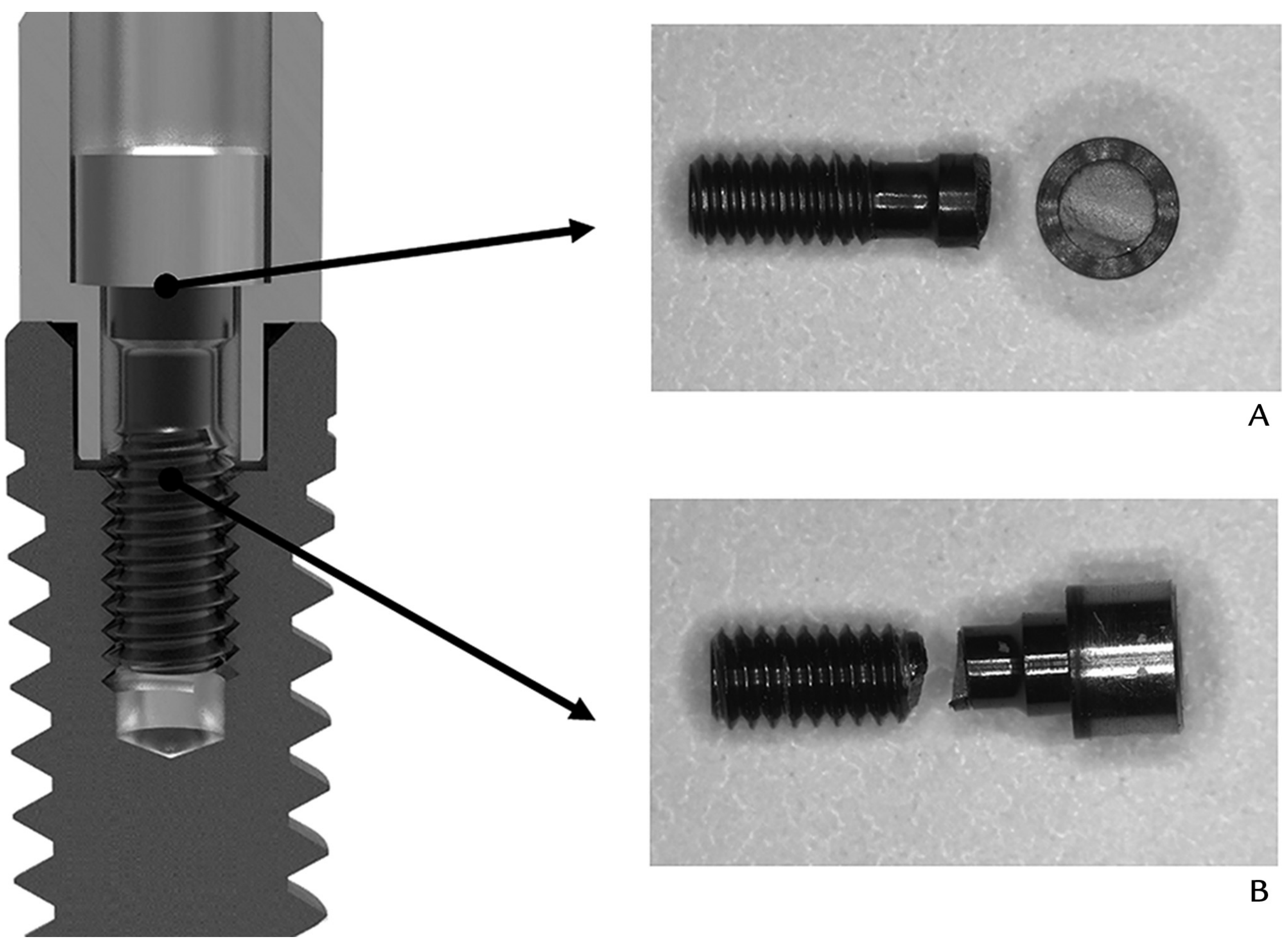

**Figure 7.** Fatigue failure. A, Head-shank transition section in roll-threaded screw. B, First engaged thread in cut-threaded screw.

**Table 5.** Maximum loads measured in static tests (N=5)

| | Maximum Static Loads (N) | |
|---|---|---|
| | Cut Thread | Rolled Thread |
| Test 1 | 676 | 602 |
| Test 2 | 622 | 656 |
| Test 3 | 568 | 705 |
| Test 4 | 600 | 593 |
| Test 5 | 689 | 704 |
| Mean | 631 | 652 |
| Standard deviation | 51 | 54 |

Limitations of this in vitro study included that a single-screw model with single rolled and cut thread manufacturing parameters were analyzed. Thus, further studies should evaluate different screw models and analyze the effect of alternative manufacturing variables.

## CONCLUSIONS

Based on the findings of this in vitro study, the following conclusions were drawn:

1. Roll-threaded prosthetic screws had better mechanical performance than cut-threaded screws.
2. Rolling placed compressive residual stresses on the thread surface, which improved the fatigue behavior of the dental implant restoration assembly without affecting its ability to withstand static load.
3. The rolling process reproduced the geometry of the rolled thread profile poorly, but this problem might be solved by altering the manufacturing parameters.

## REFERENCES

1. Warreth A, Fesharaki H, McConville R, McReynolds D. An introduction to single implant abutments. Dent Update 2013;40:7-17.
2. Patterson EA, Johns RB. Theoretical analysis of the fatigue life of fixture screws in osseointegrated dental implants. Int J Oral Maxillofac Implants 1992;7:26-33.
3. Barbosa GS, da Silva-Neto JP, Simamoto-Júnior PC, das Neves FD, da Gloria Chiarello de Mattos M, Ribeiro RF. Evaluation of screw loosening on new abutment screws and after successive tightening. Braz Dent J 2011;22:51-5.
4. Bickford JH. Introduction to the design and behavior of bolted joints. 4th ed. New York: CRC Press; 2008. p. 1-14.
5. Lang LA, Kang B, Wang RF, Lang BR. Finite element analysis to determine implant preload. J Prosthet Dent 2003;90:539-46.
6. Stüker RA, Teixeira ER, Beck JCP, Da Costa NP. Preload and torque removal evaluation of three different abutment screws for single standing implant restorations. J Appl Oral Sci 2008;16:55-8.
7. Çehreli MC, Akça K. Iplikçioğlu H, Şahin S. Dynamic fatigue resistance of implant-abutment junction in an internally notched Morse-taper oral implant: influence of abutment design. Clin Oral Implants Res 2004;15:459-65.
8. Armentia M, Abasolo M, Coria I, Bouzid AH. On the use of a simplified slip limit equation to predict screw-loosening of dental implants subjected to external cyclic loading. Appl Sci 2020;10:6748.
9. Dixon DL, Breeding LC, Sadler JP, McKay ML. Comparison of screw loosening, rotation, and deflection among three implant designs. J Prosthet Dent 1995;74:270-8.
10. Budynas R, Nisbett JK. Fatigue failure resulting from variable loading. In: Shigley's mechanical engineering design. 8th ed. New York: McGraw-Hill; 2006. p. 260-348.
11. Shemtov-Yona K, Rittel D. Fatigue of dental implants: facts and fallacies. Dent J 2016;4:1-16.
12. Stappert CFJ, Baldassarri M, Zhang Y, Hänssler F, Rekow ED, Thompson VP. Reliability and fatigue failure modes of implant-supported aluminum-oxide fixed dental prostheses. Clin Oral Implants Res 2012;23:1173-80.
13. Dhima M, Paulusova V, Lohse C, Salinas TJ, Carr AB. Practice-based evidence from 29-year outcome analysis of management of the edentulous jaw using osseointegrated dental implants. J Prosthodont 2014;23:173-81.
14. Pjetursson B, Asgeirsson A, Zwahlen M, Sailer I. Improvements in implant dentistry over the last decade: comparison of survival and complication rates in older and newer publications. Int J Oral Maxillofac Implants 2014;29:308-24.
15. Pérez MA. Life prediction of different commercial dental implants as influence by uncertainties in their fatigue material properties and loading conditions. Comput Methods Programs Biomed 2012;108:1277-86.
16. Yamanaka S, Amiya K, Saotome Y. Effects of residual stress on elastic plastic behavior of metallic glass bolts formed by cold thread rolling. J Mater Process Technol 2014;214:2593-9.
17. Armentia M, Abasolo M, Coria I, Albizuri J. Fatigue design of dental implant assemblies: a nominal stress approach. Metals (Basel) 2020;10:744.
18. Wang RR, Fenton A. Titanium for prosthodontic applications: a review of the literature. Quintessence Int 1996;27:401-8.
19. Ohkubo C, Watanabe I, Ford JP, Nakajima H, Hosoi T, Okabe T. The machinability of cast titanium and Ti-6Al-4V. Biomaterials 2000;21:421-8.
20. Domblesky JP, Feng F. A parametric study of process parameters in external thread rolling. J Mater Process Technol 2002;121:341-9.
21. Saglam H, Kus R. Performance of internal thread rolling head and the mechanical properties of rolled thread. In: Proceedings of 6th International Advanced Technologies Symposium. IATS'11 2011:210-7.
22. Zhang DW, Zhao SD, Ou H. Analysis of motion between rolling die and workpiece in thread rolling process with round dies. Mech Mach Theory 2016;105:471-94.
23. Janíček L, Petruška J, Maroš B, Rusz S. Cold forming of bolts without thermal treatment. J Mater Process Technol 2002;125:341-6.
24. Zhelezkov OS, Malakanov SA, Semashko VV. Prediction of thread rolling tools wear resistance. Procedia Eng 2017;206:630-5.
25. Furukawa A, Hagiwara M. Estimation of the residual stress on the thread root generated by thread rolling process. Mech Eng J 2015;2:14-00293.
26. Lange K. Handbook of metal forming. 1st ed. New York: McGraw-Hill; 198512.1-12.33.
27. Darshith S, Ramesh Babu K, Manjunath SS. Comprehensive study of cut and roll threads. IOSR J Mech Civ Eng 2014;11:91-6.
28. Khostikoev MZ, Mnatsakanyan VU, Temnikov VA, Maung WP. Quality control of rolled threads. Russ Eng Res 2015;35:143-6.
29. Kim W, Kawai K, Koyama H, Miyazaki D. Fatigue strength and residual stress of groove-rolled products. J Mater Process Technol 2007;194:46-51.
30. Oberg E, Jones FD, Horton HL, Ryffel HH. In: Machinery's handbook: a reference book for the mechanical engineer, designer, manufacturing engineer, draftsman, toolmaker, and machinist. 29th ed. New York: Industrial Press; 2012. p. 2048-52.
31. Stephens RI, Bradley NJ, Horn NJ, Gradman JJ, Arkema JM, Borgwardt CS. Fatigue of high strength bolts rolled before or after heat treatment with five different preload levels. SAE Tech Pap 2005.
32. Cullity BD. Elements of X-ray diffraction. 2nd edition. Prentice Hall; 1978. p. 149-232.
33. Fitzpatrick ME, Fry A, Holdway P, Kandil F, Shackleton J, Suominen L. Measurement good practice guide No. 52. Determination of residual stresses by X-ray diffraction. measurement good practice guide. Middlesex: National Physical Laboratory Teddington; 2005. p. 12-29.
34. Schajer GS. Practical residual stress measurement methods. practical residual stress measurement methods. Hoboken: John Wiley & Sons Inc; 2013. p. 139-61.
35. International Organization for Standardization. ISO 14801:2007. Dentistry. Implants. Dynamic fatigue test for endosseous dental implants. Geneva: International Organization for Standardization; 2007. (Date: 2006-11).
36. Fatigue and scatter. In: Schijve J, editor. Fatigue of structures and materials. Dordrecht: Springer Netherlands; 2009. p. 373-94.
37. Bickford JH. Introduction to the design and behavior of bolted joints. 4th ed. New York: New York: CRC Press; 2008. p. 48-54.
38. ASTM International. ASTM E739–E791. Standard practice for statistical analysis of linear or linearized stress-life (S-N) and strain-life (e-N) fatigue data. Annu B ASTM Stand 2004.

39. Cui M, Zhao S, Chen C, Zhang D, Li Y. Finite element modeling and analysis for the integration-rolling-extrusion process of spline shaft. Adv Mech Eng 2017;9:1-11.
40. Li Y, Zhao S. Study on the improvements of incremental rolling process for spline shaft with round tools based on finite element method. In: 2013 IEEE International Conference on Mechatronics and Automation. IEEE ICMA 2013:98-103.
41. Scott WW, DeHaemer MJ. Thread rolling. In: Machining. Ohio: ASM International; 1989. p. 280-95.
42. Tschaetsch H, Koth A. Metal forming practise: Processes - Machines - Tools. 1st ed. Berlin: Springer; 2006. p. 1-405.
43. Ruiz-Hervias J, Atienza JM, Elices M. Synchrotron strain scanning for residual stress measurement in cold-drawn steel rods. J Strain Anal Eng Des 2011;46:627-37.
44. Rotundo F, Korsunsky AM. Synchrotron XRD study of residual stress in a shot peened Al/SiC pcomposite. Procedia Eng 2009;1:221-4.
45. Khan MK, Fitzpatrick ME, Hainsworth SV, Evans AD, Edwards L. Application of synchrotron X-ray diffraction and nanoindentation for the determination of residual stress fields around scratches. Acta Mater 2011;59:7508-20.

**Corresponding author:**

Mr Mikel Armentia
Department of Mechanical Engineering
University of the Basque Country (UPV/EHU)
Plaza Ingeniero Torres Quevedo, 1
Bilbao, 48013
SPAIN
Email: marmentia002@ikasle.ehu.eus

**Acknowledgments**

The authors thank Aitor Larrañaga (SGIker, Bizkaia, Spain) and Raúl Cosgaya for the valuable assistance provided in the residual stresses measurements and the fatigue tests performed in this work, respectively.